\documentclass[aps,prl,twocolumn,groupedaddress]{revtex4}

\usepackage{graphicx, hyperref, acronym}
\usepackage{amssymb, amsmath}
\usepackage{epstopdf}

\newcommand{\ra}{\rangle}

\newcommand{\eq}[1]{\begin{equation} #1 \end{equation}} 

\newcommand{\A}{\mathcal{A}}
\newcommand{\T}{\mathcal{T}}
\newcommand{\U}{\mathcal{U}}
\newcommand{\C}{\mathcal{C}}

\newcommand{\ket}[1]{{\left| #1 \right\rangle}}

\newcommand{\hC}{\hat{\chi}}
\newcommand{\hN}{\hat{N}}

\newcommand*\xbar[1]{%
  \hbox{%
    \vbox{%
      \hrule height 0.5pt 
      \kern0.3ex
      \hbox{%
        \kern-0.1em
        \ensuremath{#1}%
        \kern-0.1em
      }%
    }%
  }%
} 

\newcommand{\V}{\xbar{V}}

\acrodef{CFT}{conformal field theory}
\acrodef{FQH}{fractional quantum Hall}
\acrodef{TSB}{topological symmetry breaking}

\begin{document}


\title{Topological symmetry breaking: Domain walls and partial instability of chiral edges}


\author{F.A. Bais}
\email[]{F.A.Bais@uva.nl}
\affiliation{Institute for Theoretical Physics, University of Amsterdam, Science Park 904, 1090 GL Amsterdam, The Netherlands}
\affiliation{Santa Fe Institute, Santa Fe, New Mexico 87501, USA}
\author{S.M. Haaker}
\email[]{S.M.Haaker@uva.nl}
\affiliation{Institute for Theoretical Physics, University of Amsterdam, Science Park 904, 1090 GL Amsterdam, The Netherlands}


\date{\today}

\begin{abstract}
In 2D topological systems chiral edges may exhibit a spectral change due to the formation of a Bose condensate and partial confinement in the bulk according to the topological symmetry breaking (TSB) mechanism.
We analyze  in detail what happens in the bulk as well as on the edge for a set of simple chiral fractional quantum Hall systems. TSB corresponds to the spontaneous breaking of a global discrete symmetry and therefore to the appearance of domain walls in the bulk and kinks on the edge. The walls are locally metastable and  break if a confined particle-hole pair  is created. 
\end{abstract}


\maketitle
Topologically ordered phases of matter exhibit many fascinating properties, such as fractionalization of spin and charge and the possibility of non-Abelian braiding statistics 
\cite{Nayak2008}. One phenomenon that plays a pivotal role in almost all topological phases --- be they topological insulators, topological superconductors, fractional quantum Hall (FQH) states {\em et cetera} --- is the appearance of protected edge modes. In fact, for phases of which the bulk can be described by a Chern-Simons (CS) theory there is a bulk-boundary correspondence stating that the (1+1)-dimensional gapless edge can be described by a conformal field theory (CFT) and the bulk wave functions by the conformal blocks of the same CFT \cite{Moore1991,Wen1995}. This strongly suggests that as long as the bulk is topologically ordered, no perturbations can destroy the chiral gapless edge theory. For nonchiral edges there is the possibility of counter-propagating edge modes gapping out and a criterium for stable edges is given in in terms of the Lagrange subgroup in \cite{Haldane1995,Levin2013}, which has also been formulated  for symmetry enriched phases in \cite{Lu2012, Hung2013, Hung2013b, Gu2014}.
In this Letter, we extend this picture. We show that a careful treatment of the problem necessarily has to take into account the possibility of Bose condensation in the bulk, corresponding to  topological symmetry breaking (TSB), which refers to phase transitions between different topologically ordered phases due to the condensation of bosonic quasiparticles breaking the quantum group symmetry \cite{Bais2002, Bais2003, Bais2009}. This formalism has been successfully applied to many transitions between topological phases \cite{Barkeshli2010,Barkeshli2011, Bais2009b,Bais2014} and a more mathematical treatment can be found in \cite{Kong2013, Kitaev2012}.  We will show how chiral models corresponding to specific Laughlin states, the simplest of these occur at $\nu=\frac{1}{8}$ and $\nu=\frac{1}{9}$, may become unstable due to TSB and will give a detailed description of possible domain walls and confined particles in the broken phase. 

We will recall the main ingredients of TSB here and refer to the literature for more details.  
TSB involves three steps: i) We start from a (2+1)-D topologically ordered phase with topological excitations that fall into irreps of some quantum group $\A$. The different sectors $a,b,\ldots \in \A$ carry topological quantum numbers related to their interaction under fusion and braiding, such as quantum dimension and topological spin. ii) Whenever there is a bosonic sector present in this phase, it may condense and form a new ground state where charge is only defined up to the charge of the condensate. Sectors of $\A$ may become identified with each other or split up forming new sectors, we denote the  resulting intermediate phase by $\T$. The new sectors have well-defined fusion rules, but braiding might  be inconsistent. iii) The requirement that all sectors should have trivial monodromy with the condensate implies that those who do not, become confined in the bulk and are expelled to the edge of the sample. The remaining sectors, having well-defined braiding and fusion, form the final phase with quantum group $\U\subset \A$. 



To be definite and explicit we will consider the specific example of a Laughlin state at $\nu=1/8$ \endnote{In the supplementary material we will show how to generalize this to Laughlin states at other filling fractions, including fermionic states.}. The bulk is described by a $U(1)$ Chern-Simons field at level $k=8$ and the edge has gapless edge modes corresponding to a chiral boson theory compactified at radius $R=\sqrt{8}$. In the bulk the CS field can be written as a pure gauge $a_i(z)=\partial_i\phi(z)$, but on the edge $\phi$ becomes dynamical with Lagrangian \cite{Wen1995}
\begin{equation}
\mathcal{L}=\frac{1}{4\pi}\int\left( \partial_x\phi\partial_t\phi-\partial_x\phi\partial_x\phi\right) ,
\label{eq:lagrangian}
\end{equation}
where $x$ is the coordinate along the edge.
It is a chiral boson with a global $U(1)$ invariance corresponding to the transformation $\phi(x)  \rightarrow \phi(x) + f$. 
The mode expansion of $\phi(x)$ compactified on a radius $R=\sqrt{8}$ on a cylinder of circumference $L$ is
\eq{
\phi(x)= \frac{2\pi \hN}{\sqrt{8}L}\;x + \sqrt{8}\;\hC +\text{oscillator modes},
}
where we will discard the oscillator modes as we are only interested in the distinct topological sectors. The conserved current $\partial_x\phi$ leads to a conserved charge $\frac{1}{2\pi R}\int_0^L \partial_x\phi dx=\hat{N}/R^2$, which commutes with the zero mode as $[\hC, \hN]=i$.

Let us introduce some operators that play a crucial role in the present TSB analysis. The operators that create a localized topological charge are the normal ordered vertex operators 
\eq{
V_n(x)=:\!e^{i\frac{n}{\sqrt{8}}\phi(x)}\!: \ ,
}
where we will omit writing the normal-ordering symbol in the rest of this Letter. The vertex operators
 have conformal weight $h_n=\frac{n^2}{16}$ and transform as irreps under the   global symmetry $\phi(x)\rightarrow \phi(x)+f$: $V_n\rightarrow e^{inf/R}V_n$. 
The operators that measure charge correspond to a nonlocal Wilson loop operator, defined as
\eq{ \label{eq.wilsondef}
W_q=\exp\Big[\frac{i q}{\sqrt{8}}\int_{-L/2}^{L/2} a_x dx\Big]=e^{2\pi i q \hN/8}.
}
It is the exponentiated conserved global charge operator which is invariant under the global $U(1)$ symmetry. Note that this operator can  be extended to the bulk where it becomes locally gauge invariant 
  for any value of $q$ and we will use it  at various values of $q$ to probe the phase structure of the theory later on.
We will also employ open Wilson line operators
\eq{\label{eq:wilsonline}
W_q(x_1, x_2)=\exp\Big[\frac{iq}{\sqrt{8}}\int_{x_1}^{x_2} \partial_x\phi(x) dx\Big]\;,
}
which are still invariant under the global $U(1)$ symmetry, but if  extended to the bulk are not invariant under the local $U(1)$ which takes $\phi(z)\rightarrow \phi(z)+f(z)$. The  gauge invariant operator which is  well defined everywhere is obtained  by attaching  charged quasiparticles represented by vertex operators to the integer charged Wilson lines 
\eq{\label{eq:bulkline}
V_n(z_1)W_n(z_1, z_2)V^\dagger_n(z_2).
}
Gauge invariance  imposes that anyons and anti-anyons come in pairs  connected by a Wilson line, which in the present case is just a Dirac string, a gauge artifact  which can be moved around without changing the physics. This reflects the bulk-boundary correspondence, if we want to insert a single quasiparticle in the bulk there has to be an antiparticle somewhere on the boundary too.

Let us return to the edge theory. We label the distinct topological charge sectors by $n=0,\ldots,7$, which is the eigenvalue of $\hN$,
and they can be created by acting with the nonlocal operator
\eq{
\V_n=\exp\Big[i\frac{n}{\sqrt{8}L}\int_{-L/2}^{L/2}\phi(x)dx \Big]=e^{in\hC}.
}
on the vacuum state  $\V_n\ket{0}=\ket{n}$. One also finds that
$[ \hN, \V_n]=n\V_n$, so they act  as ladder operators on the charge eigenstates. 
The 8 topological sectors form irreps of a global $Z_8$ group generated by $ W_1=  e^{\frac{2\pi i}{8} \hat{N}}$ ($W_1^8=1$) and the ground state $|0\ra$ is unique. 
The algebra of the operators that create a unit of charge (both $V_n(x)$ and $\V_n$) and measure charge is
\begin{equation}
W_m V_n = V_nW_m e^{ 2\pi n m  i/8}. 
\label{eq:VWalgebra}
\end{equation}
 Let us briefly show how TSB applies to this specific situation.
 The topological excitations of the chiral boson theory at $R=\sqrt{8}$ are described by the quantum group $\A=U(1)_4$. When we label them by $n=0,\ldots, 7$, their spin and quantum dimension are given by $h_n=n^2/16$ and $d_n=1$, respectively. There is one nontrivial boson, $n_b=4$ and when it condenses we have to identify $n\sim n+4$ in the broken phase $\T$, which has fusion rules $Z_4$. The odd sectors have nontrivial monodromy with $n_b$, therefore are confined and  expelled from the bulk. This leaves us with a broken unconfined phase $\U=U(1)_1$, corresponding to a compactified boson at $R=\sqrt{2}$, which has one nontrivial sector with spin $h=\frac{1}{4}$. Even though there is no backscattering in $\A$ because of its chirality, we claim that the edge current is not entirely protected as particles can disappear into the condensate. Our goal in this Letter is to present a very physical, field theoretical interpretation of  these phases. We will see that different domains may form and show what the precise role of the confined particles is.

To probe the phase structure we use some of the vertex operators  as order parameters  that may or may not acquire a finite expectation value. 
Clearly, in the unbroken phase $\A$  all have vanishing vacuum expectation value, but in the broken we will assume that for a $V_k$ we have
\begin{equation}
|\langle V_k \rangle|^2 \neq 0.
\end{equation}
This means that the ground state should be of the form $|n\ra_k=\sum_{l}c_l|n+lk\ra$. If in addition we require this new ground state to be invariant under rotations by $2\pi$ generated by $\hat{R}= e^{i2\pi \hat{N}^2 / 16}$ it will further restrict the state to
$n=0$ or $n=4$ and for both of these values of the charge, we need to set $k=4$. So in this new phase the two possible ground states are 
\begin{equation}
\ket{ 0}_\pm= \frac{1}{\sqrt{2}}\big(|0\ra \pm |4\ra\big)
\end{equation}
The condition that the ground state is invariant under $\hat{R}$ is of course equivalent to demanding integer spin which is a property a condensable sector should have. As a result, a careful treatment   of TSB shows  that we have  two degenerate  ground states instead of only one, which may result in the formation of different domains  due to spontaneous symmetry breaking. 
The operator $V_4$ has nonzero expectation value which is very similar to creating Cooper pairs in a superconductor. In our case, we can freely create and annihilate particles of topological charge $n=4$. The states rearrange themselves as eigenstates of $V_4$
\begin{equation}
\label{eq:Tsectors}
 |n\ra_\pm =  \frac{1}{\sqrt{2}}\left(|n\ra \pm |n+4\ra\right),\qquad  n=0,\ldots,3 \ .
\end{equation}
They form an orthonormal and a complete set of representations of the group $G=Z_4\otimes Z_2$, generated by $W_2=e^{2\pi i \hat{N}/4}$ and $V_4$. These two operators commute, $[W_2,V_4]=0$ and the action on the states is
\begin{eqnarray}
\label{ }
W_2\; |n\ra_\pm &=& e^{2\pi i n/4}  |n\ra_\pm  \\
V_4\; |n\ra_\pm &=&\pm  |n\ra_\pm 
\end{eqnarray}
In the  broken phase the ground state is twofold degenerate
and the two states are mapped onto each other by $W_1$
\begin{equation}
W_1\;|0\rangle_\pm=| 0\rangle_\mp .
\end{equation}
They carry a $Z_2$ charge generated by $V_4$, and  the trivial  $Z_4$ charge under $W_2$. Due to the twofold vacuum degeneracy the system behaves similarly to the 2D Ising model without external field in the ordered phase. 

Let us focus again on the edge and start from a state where the entire edge is in the ground state $|0\ra_+$. We  apply the  open Wilson line operator in Eq. \eqref{eq:wilsonline}
which creates a domain  in the $|0\ra_-$ phase in between the points $x_1$ and $x_2$
and consequently also a kink anti-kink pair in $\phi$ at $x_1$ and $x_2$.
The localized kinks  have finite energy, so in the edge theory the kinks are just massive solitons which are not confined because there is only vacuum in between them. Note that on the edge we only have the global symmetry of shifting $\phi(x)$ by a constant, clearly the Wilson line on the edge is invariant under this transformation and creating these different domains does not require the introduction of the $V_1$ sectors. To understand this we should extend our discussion to include the bulk which is what we do next.

\begin{figure}[htbp]
\includegraphics[width= 0.1\textwidth ]{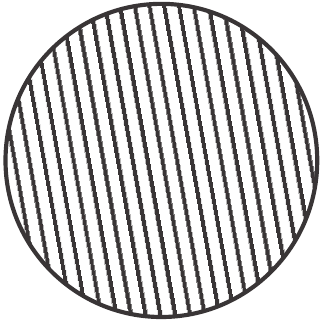}\hspace{.2cm}
\includegraphics[width= 0.1\textwidth ]{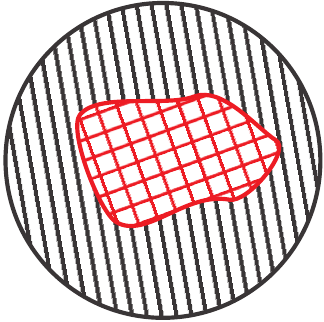}\hspace{.2cm}
\includegraphics[width=0.1\textwidth ]{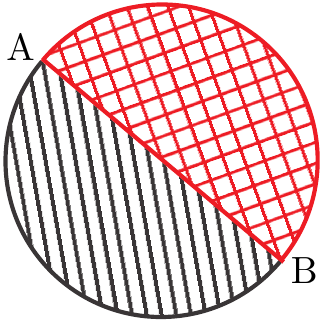}\hspace{.2cm}
\includegraphics[width=0.1\textwidth ]{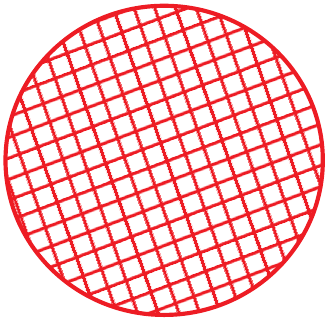}
\caption{\small A disc in the broken phase $\U$ with two possible vacua: $+$ vacuum is indicated by gray stripes, and $-$ vacuum  indicated by a red  mesh. On the left we start with the disc entirely in the $+$ vacuum. In the next figure we have created a Wilson loop $W_1$ along a contour $\mathcal{C}$, which creates a $-$ vacuum inside the loop and a physical domain wall along the loop. 
It can be deformed to lie partially on the boundary as indicated in the third figure. The boundary now has two different domains where the kinks located around point ${\rm A}$ and ${\rm B}$ carry energy and they are connected to the wall going across the disc through the bulk, which also carries energy. In the last figure the entire Wilson loop lies on the boundary and the entire disc is in the $-$ vacuum state.
}
\label{fig:wilsonloops}
\end{figure}

We may extend the operators $W_q(\C)$ and $V_n(z)$ to  well-defined operators referring to closed loops $\C$ and points (punctures) $z$  in the bulk.  In discussing the phase structure of the broken phase $\U$, it is important to make a clear distinction between the cases where we do or do not allow  insertions of the confined sector $V_1(z)$ in the  bulk. 
This distinction can be made because there are two scales in the problem: the gap or mass of the $V_n$ excitations and the presumably smaller energy scale associated with the condensate.
It is most natural to start with a situation where we do not include them, but
we may still consider the Wilson loop operators with arbitrary $q$ and in particular also with $q=1$. 
The interpretation of closed loops is similar to that on boundary, the Wilson loop operator now creates a domain of -- vacuum in the bulk, and a domain wall along the contour $\C$.  It is interesting to deform this configuration as indicated in Fig. \ref{fig:wilsonloops}.
In the  figures on the left we have sketched the situation just discussed. But the loop can be moved around at will, in particular we may put it partially along the edge as in the third figure. How to interpret this physically? 
When looked at  from the perspective of the boundary we see that at the points ${\rm A}$ and ${\rm B}$  where the closed loop leaves the boundary, the vacuum flips and therefore there should be a kink in the field on the boundary. The other part of the contour, going from ${\rm B}$ to ${\rm A}$ through the bulk, is a massive domain wall ending at the kink-antikink pair. 
The  situation is comparable to  the states created by $V_n(z)$ which represent massive localized anyons in the bulk and massless modes on the edge. 


When there are different domains in the bulk we have to probe the system locally to measure which domain we are in, we cannot simply use $V_4(z)$ as it is not gauge invariant. Instead we need to use the gauge invariant object in Eq. (\ref{eq:bulkline}) with $n=4$, which gives $+ 1$ if the end points are in the same vacuum and $-1$ if they are in different vacua, i.e the Wilson line crossing the wall either an even or odd number of times. The coloring of domains in the pictures of Fig. \ref{fig:wilsonloops} is therefore well-defined and in the restricted setting without  $V_1$ excitations the vacuum states can be unambiguously carried over to the bulk. 
This establishes a detailed, consistent  picture of the physics of TSB, except that we still have  to consider the role of the vertex operator $V_1(z)$ in the bulk of $\U$. 
\begin{figure}[t!]
\begin{center}
\includegraphics[width= 0.1\textwidth ]{disc3.pdf}\hspace{1cm}
\includegraphics[width= 0.1\textwidth ]{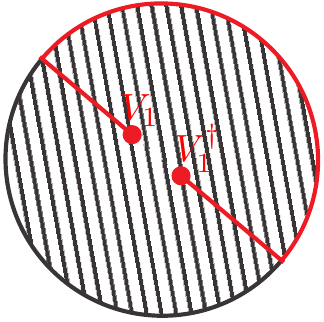}
\caption{\small In the left figure  a closed $W_1$ loop is depicted and  on the right the loop is broken through the creation of a $V_1$ pair. Probing with a $W_4$ Wilson line to a point on the boundary is still consistently and unambiguously defined. It makes no difference  whether this line crosses the wall or passes through the opening. In the one case the phase jump occurs because of the crossing, and in the other it comes from braiding with the $V_1$ vertex. However, it is also clear that the coloring of the domains in the bulk does not make any sense if we allow the $V_1$ anyons to be produced,  but restricting one self to only the edge the coloring still does make sense.}
\label{fig:openloops}
\end{center}
\end{figure}

Next we will consider the stability of  a domain  wall created by $W_1(\mathcal{C})$. We have to distinguish two types of instabilities. One is a {\it global} instability meaning that a closed loop in the bulk can shrink to zero;  because the wall has a fixed  energy per unit length, shrinking lowers the total energy of the configuration and there is no topological obstruction to fully contract. 
More interesting  is  a {\it local}  topological instability of the wall. The wall can in principle break into the Wilson line Eq. \eqref{eq:wilsonline} upon the  creation of a $V_1$ particle-hole pair attached to the new endpoints of the broken wall, this process is depicted in Fig. \ref{fig:openloops}. Whereas the $V_1$ quasiparticles are not confined on the boundary as we argued before, they are linearly confined in the bulk exactly because they have to be attached to a domain wall with finite energy per unit length. The walls are metastable  because the creation of a massive pair requires  an energy of at least twice the particle gap. Another important consequence of the metastability of the wall in the bulk, is that the domain structure of the vacuum is no longer protected.

A topological argument explaining this goes as follows \cite{Bais1981}. In the unbroken phase $V_1$ is  present  and we have a full $U(1)/Z_8$ gauge group in the bulk with topological flux/particle sectors  $\pi_1(U(1)/Z_8) = \pi_0(Z_8) = Z_8$, corresponding to representations of the $Z_8$ group generated by $W_1$. After breaking, the gauge group is formally changed to $U(1)/Z_4$ corresponding to the $Z_4$ subgroup of $Z_8$ consisting of the even  elements, and $\pi_0(Z_4)$ refers to the even sectors  $V_{2n}$.  The homotopy sequence of interest here is
\begin{equation}
\label{eq:sequence}
\pi_0(Z_4) \rightarrow \pi_0(Z_8) \rightarrow \pi_0(Z_8/Z_4)\;,
\end{equation}
implying that the Image of the first mapping is the Kernel of the second. In physical terms this means that the  even sectors of $\pi_0(Z_8)$ get mapped onto the trivial sector of $\pi_0(Z_8/Z_4) = Z_2$, where the latter group  labels per definition the new types of domain wall that arise in the broken phase. In other words, the odd charges of the $\pi_0(Z_8)$ are mapped onto the nontrivial domain walls and are therefore confined, exactly as advertised.

  


The overall picture remains completely  consistent if one takes into account that now there are two ways to go from a $+$ state created by $(V_n(x)+V_{n+4}(x))|0\rangle$ at a position $x$ on the edge to the left of point A in Fig. \ref{fig:openloops}, to the corresponding $-$ state with $x$  on the right.  The first option is to `cross' the wall by transforming a vertex operator with $W_1$ using (\ref{eq:VWalgebra}). The other is by moving the vertex operators involved through the bulk and around the endpoint at the opening in the wall, which in fact  means  acting with the monodromy operator. 
Let us demonstrate this explicitly by considering this question in the original unbroken $\A$ theory and see what can be carried over to the broken phase.

Given that the monodromy, i.e. encircling an anyon $V_n$ with $V_m$ (in the original $\A$-theory, with $n, m=0,\ldots,7$), yields a phase factor 
\begin{equation}
\label{eq:monodromy }
\theta_{mn} = \exp (2\pi i (h_{m+n}- h_n - h_m  )) =  e^{2\pi i nm/8} \ ,
\end{equation}
we can make some important observations. Since the $\A$ sectors become combined (identified) as in Eq. \ref{eq:Tsectors} in the broken phase, encircling them around another $\T$ sector gives different monodromy phases.
For $n,m=0,\ldots 3$ and  $k,k'= 0,1$, the different phases of the monodromy can be expressed as
\begin{align}
\label{ }
 &\frac{2\pi}{16}\left\{  (n+m+4k+4k')^2-(n+4k)^2 - (m+4k')^2\right\} =\nonumber\\ 
&=\frac{2\pi}{8}(nm +4nk' +4mk)\; \mod 2\pi
\end{align}
Two $\T$ sectors have consistent braiding if their monodromy is independent of $k$ and $k'$, which leaves us with only the sectors $n=0$  and $n=2$, as expected. One also may verify that only these are mutually local with respect to the new vacuum and therefore survive as unconfined sectors. 

Note that from the monodromy phase  we learn  that if we have the  fundamental quasiparticle corresponding to $V_1$ in the bulk,  and bring the new vacuum $V_0\pm V_4$ around it, that would map the two vacua onto each other, i.e. $(V_0 \pm  V_4) \rightarrow  (V_0\mp V_4)$. 
This means that the net effect of moving around the confined particle is the same as crossing the wall as we have described above, where the $\pm$ state  transforms under $W_1$ and gets mapped to the $\mp$ state at the other side of the wall.
This shows once more that the wall is not locally stable, and it can break  under the creation of a fundamental quasiparticle/hole pair, each of them remaining attached to the newly created  end points. Alternatively, if one starts from a disc entirely in one of the vacua and creates a $V_1$-$V_1^\dagger$ pair in the bulk,  then both particles will stay connected by a wall meaning that they are confined indeed. So their appearance will be exponentially suppressed not only because of their mass but also because of their interaction energy that rises linearly with distance.


In this Letter we mainly focused on $\A=U(1)_4$, but the construction is easily generalized to other Laughlin states at lower filling fraction. The details of these states can be found in the supplementary material as well as comments on how the domain walls and confinement generalizes.
This work suggest many interesting generalizations such as analyzing the field theory of TSB in general non-Abelian theories with a WZW theory  on the edge.

We conclude by summarizing our results. Even though, there can be no backscattering in a chiral system, we have shown that certain chiral edges are not entirely protected because TSB may occur. Certain topological sectors form a condensate and others become confined in the bulk and acquire a mass on the edge. Furthermore, we have extended our understanding of the original TSB picture proposed in 2002, by introducing an explicit order parameter which obtains a finite expectation value in the broken phase. We identified  degenerate ground states leading to different domains separated by domain walls, giving a nice understanding of the confined particles in the bulk. Remarkably, these particles  turn out not to be confined on the edge but do acquire a mass there. We stated simple criteria for the stability of the domain walls. This work  also clearly shows the essential observable differences between an exact $\U$ theory and the $\U$ phase obtained after TSB of an $\A$ theory. \\

\begin{acknowledgments} The authors thank Jesper Romers, Sebas Eli\"ens
for their contributions at an earlier stage of this work. The work of S.M.H. is funded by the Foundation for Fundamental Research on Matter (FOM), which is part of the Netherlands Organisation for Scientific Research (NWO).
\end{acknowledgments}

\hfill


\pagebreak
\section{Supplementary material}\label{sup}

%

We will derive for which Laughlin states at filling fraction $\nu=1/M$ the particle spectrum contains a nontrivial boson, which could drive a phase transition to a state at different filling fraction. As explained in the main part of this Letter, the edge as well as the bulk can be described by a chiral boson with compactification radius $R=\sqrt{M}$. It is well known that for rational radius, i.e.
\eq{
R=\sqrt{\frac{2p'}{p}}\ , \quad p, p' \text{~are coprime}\ ,\label{radius}
}
there are only a finite number of sectors, because the vertex operator $\Gamma^\pm=e^{\pm i\sqrt{2pp'}\phi}$ can be added to the current algebra. 
The primary fields of the extended theory must have local OPE with the currents and are of the form
\eq{
V_n=e^{in\phi/\sqrt{2pp'}} \ .
}
Their weights are
\eq{
h_n=\frac{n^2}{4pp'}\ , \qquad n=0,\ldots, 2pp'-1 \ ,
}
and the Hilbert space falls into irreps of the extended algebra which will be denoted by $U(1)_{pp'}$. Note that all of the above is invariant under the interchange $p\leftrightarrow p'$, which corresponds to the invariance under modular transformations.


\subsection{Bosonic Laughlin states}\label{boslaughlin}

We first turn to the bosonic Laughlin states at filling fraction $\nu=1/M$, with $M$ even. The theory can be described by a compactified chiral boson at radius $R=\sqrt{M}$ where we add the bosonic operator $V_M=e^{i\sqrt{M}\phi}$ to the chiral algebra, resulting in a finite number of sectors which form an algebra denoted by $U(1)_{M/2}$. The theory we start from has $M$ sectors $V_n=e^{in\phi/\sqrt{M}}$, with conformal weights (spins)
\eq{\label{spinboslaughlin}
h_n=\frac{n^2}{2M}\ , \qquad n=0,...,M-1\ .
}
The vertex operator that describes the physical boson of charge $e$ is precisely the  operator that is added to the chiral algebra $V_e=V_M$, with weight $h_e=M/2$, which is an integer for even $M$.

\subsection{Fermionic Laughlin states}\label{ferlaughlin}


 Moore and Read treated the fermionic case in \cite{Moore1991}. They choose the electron operator $V_e=e^{i\sqrt{M}\phi}$ as extended generator  even though it has half integer spin.  The other operators are chosen  such that they are mutually local with the electron operator. The weights of these operators are $h_n=\frac{n^2}{2M}$, with $n=0,\ldots,M-1$ and they indeed carry the right quantum numbers associated with the quasiholes of the FQH state. The reason why we do not adopt their  formulation is that they  do not distinguish between the electron operator and the trivial operator, therefore we would never be able to distinguish between a fully gapped (edge) system and a $\nu=1$ state.


We will follow a different strategy.  For $\nu=1/M$, with $M$ odd we start from a compactified boson at $R=\sqrt{M}$. Since $M$ is odd we can choose $p'=M$ and $p=2$ as coprime integers, resulting in a $U(1)_{2M}$ theory. The weights are 
\eq{
h_n=\frac{n^2}{8M}\ , \qquad n=0,...,4M-1\ .
}
The sector with $n=2M$ corresponds to the electron and it has spin $h_{2M}=M/2$. We want all the sectors to be local with respect to the electron operator, the monodromy is given by
\eq{
{\rm M}_{n,e}=\frac{4Mn}{8M}=-\frac{n}{2} \ ,
}
which means that only the even sectors are local and are good operators in this theory. Rewriting $2m=n$, we are left with $2M$ sectors $V_m=e^{im\phi/\sqrt{M}}$, labeled by $m=0,...,2M-1$, with weights $h_m=\frac{m^2}{2M}$. Let us call this theory $U^+(1)_{2M}$, where the + denotes the even sectors. The only difference with the literature is that we count up until twice the electron. 

\subsection{Unstable bosonic Laughlin states}

We now turn to investigating which states have a nontrivial boson in its spectrum.
Starting from a phase $\A=U(1)_{M/2}$, which has $M$ sectors with spins given in Eq. \eqref{spinboslaughlin}
we will show that phases with filling fraction 
\eq{
\nu=\frac{1}{M}=\frac{1}{2l^2k}\ , \qquad l=2,3,\ldots \ ,\quad k=1,2,\ldots \ ,
}
have at least one nontrivial boson that can drive a transition to a broken phase carrying less sectors. 
 The initial phase is $\A=U(1)_{l^2 k}$, 
 corresponding to a chiral boson compactified at $R=l\sqrt{2k}$, which has $2l^2k$ sectors with spins 
 \eq{
 h_n=\frac{n^2}{4l^2k}\ , \qquad n=0,1,\ldots,2l^2k-1 \ .
 }
The smallest  nontrivial bosonic sector is $b=2lk$, which has spin $h_b=k$. When this boson forms a condensate, the other sectors arrange in orbits of length $l$ under fusion with the boson
\eq{\label{eq:orbits}
n\sim n+2lk \sim \ldots \sim n+2lk(l-1)\ ,
} 
 which means that the sectors belonging to the same orbit get identified with each other, resulting in an intermediate phase $\T$ which has fusion rules $Z_{2lk}$. 
 To check which sectors become confined we consider the monodromy of a sector $n$ with the boson
 \eq{
 {\rm M}_{n,b}=h_{n+b}-h_n-h_b=\frac{n}{l} \ .
 }
 The unconfined sectors can be expressed as $n=ml$ where $m=0,1,\ldots, 2k-1$ and we see that they have spin $h_m=\frac{m^2}{4k}$, which we recognize as a $\U=U(1)_k$ theory. Of course if this $k$ is again of the form $k=l^2k'$ there will be other bosons left in the theory, which can also condense. The highest filling fractions that are not stable are $\nu=1/8, 1/16, 1/18, \ldots$. 


\subsection{Unstable fermionic Laughlin states}

The same analysis can be performed for the fermionic Laughlin states 
and we will show that for filling fraction 
\eq{
\nu=\frac{1}{M}=\frac{1}{l^2k}, \qquad l=3,5,\ldots \ , \quad k=1,3,\ldots \ ,
}
a phase transition can occur. Starting from $\A=U^+(1)_{2l^2k}$ corresponding to a chiral boson compactified at radius $R=l\sqrt{k}$, there are $2l^2k$ sectors with spins
\eq{
h_n=\frac{n^2}{2l^2k}\ , \qquad n=0,1,\ldots, 2l^2k-1\ .
}
The charge $e$ fermion is given by $n_e=l^2k$ and has spin $h_e=l^2k/2$. 
There is a nontrivial boson in this theory  $b=2lk$, which has spin $h_b=2k$. When these particles  condense, the original $\A$ sectors rearrange into orbits of length $l$, similar to Eq. \ref{eq:orbits}. The broken intermediate phase $\T$ has fusion rules $Z_{2lk}$ and the monodromy of these sectors with the bosonic particle is
\eq{
{\rm M}_{n,b}=\frac{2n}{l}\ .
}
This demonstrates that the unconfined particles are those for which $n=lm$ with $m=0,1,\ldots,2k-1$ and their spins are given by
$h_m=\frac{m^2}{2k}$. They form a broken unconfined phase $\U=U^+(1)_{2k}$ at $R=\sqrt{k}$, corresponding to a fermionic Laughlin state at filling fraction $\nu=1/k$. The highest fractions are $\nu=1/9, 1/25, 1/27, \ldots$. For instance the $\nu=1/9$ breaks to a Laughlin state at $\nu=1$, which is an IQH phase.

\subsection{Domain walls and confinement}

It is now straightforward to generalize the discussion of domain walls and confined particles  applied to $U(1)_4$ in the main part of this Letter, to these general Laughlin states. For both the fermionic as well as the bosonic cases, the vertex operator $V_{2lk}$ acquires a vacuum expectation value, resulting in $l$ different vacua. Due to spontaneous symmetry breaking there are $l$ different possible domains in the broken phase.  There are $l-1$  Wilson loops as in Eq. \eqref{eq.wilsondef}, that create physical domain walls between these different domains and there are $l-1$ confined particles corresponding to $V_n$, with $n=1,2,\ldots, l-1$, that can break up such a Wilson loop creating a local instability as was depicted in Fig. \ref{fig:openloops}.

\end{document}